\documentclass[aps,prxquantum,reprint,twocolumn,superscriptaddress,floatfix,nofootinbib,longbibliography]{revtex4-1}

\usepackage{graphicx,amsmath,amsfonts,amssymb,amsthm,xr}
\usepackage{epsfig,amsmath,amssymb,color,dsfont,upgreek,physics}
\usepackage{mathrsfs}
\usepackage{mathtools}
\usepackage{bbold}
\usepackage{comment}
\usepackage{float}

\usepackage[bookmarks=true,colorlinks,linkcolor=OrangeRed,urlcolor=NavyBlue,citecolor=RoyalBlue]{hyperref}
\usepackage[dvipsnames]{xcolor}
\usepackage{physics}
\usepackage{amssymb}
\usepackage{amsmath}
\usepackage{amsfonts}
\usepackage{natbib}
\usepackage{hyperref}
\usepackage{graphicx}
\usepackage[dvipsnames]{xcolor}
\usepackage{bm}
\usepackage{qcircuit}
\usepackage[normalem]{ulem}

\definecolor{mygold}{rgb}{0.93,0.69,0.13}

\definecolor{mypurple}{rgb}{0.49,0.18,0.56}

\newcommand{\hc}{\mathop{\rm H.c.}}

\def\be{\begin{equation}}
\def\ee{\end{equation}}

\begin{document}

\title{Disorder-Free Localization for Benchmarking Quantum Computers}

\author{Jad C.~Halimeh${}^{*}$}
\affiliation{Max Planck Institute of Quantum Optics, 85748 Garching, Germany}
\affiliation{Department of Physics and Arnold Sommerfeld Center for Theoretical Physics (ASC), Ludwig Maximilian University of Munich, 80333 Munich, Germany}
\affiliation{Munich Center for Quantum Science and Technology (MCQST), 80799 Munich, Germany}

\author{Uliana E.~Khodaeva${}^{*}$}
\affiliation{Department of Physics, TQM Technical University of Munich, 85748 Garching, Germany}

\author{Dmitry L.~Kovrizhin}
\affiliation{LPTM, CY Cergy Paris Universite, UMR CNRS 8089, Pontoise 95032 Cergy-Pontoise Cedex, France}

\author{Roderich Moessner}
\affiliation{Max-Planck-Institut f\"ur Physik komplexer Systeme, 01187 Dresden, Germany}

\author{Johannes Knolle}
\affiliation{Department of Physics, TQM Technical University of Munich, 85748 Garching, Germany}
\affiliation{Munich Center for Quantum Science and Technology (MCQST), 80799 Munich, Germany}
\affiliation{Blackett Laboratory, Imperial College London, London SW7 2AZ, United Kingdom}

\def\thefootnote{*}\footnotetext{These authors contributed equally to this work.}

\begin{abstract}
Disorder-free localization (DFL) is a phenomenon as striking as it appears to be simple: a translationally invariant state evolving under a disorder-free Hamiltonian failing to thermalize. It is predicted to occur in a number of quantum systems exhibiting emergent or native \emph{local} symmetries. These include models of lattice gauge theories and, perhaps most simply, some two-component spin chains. 
Though well-established  analytically for special soluble examples, numerical studies of generic systems have proven difficult. Moreover, the required local symmetries are a challenge for any experimental realization. Here, we show how a canonical model of DFL can be efficiently implemented on gate-based quantum computers, which relies on our efficient encoding of three-qubit gates. We show that the simultaneous  observation of the absence of correlation spreading and tunable entanglement growth to a volume law provides an ideal testbed for benchmarking the capabilities of quantum computers. In particular, the availability of a soluble limit allows for a rigorous prediction of emergent localization length scales and tunable time scales for the volume law entanglement growth, which are ideal for testing capabilities of scalable quantum computers.

\end{abstract}

\date{\today} 

\maketitle


\section{Introduction and relevance}
More than 65 years ago, P.~W.~Anderson showed~\cite{Anderson1958} that electrons can become localized in the presence of disorder. Over the following decades, the phenomenon of localization, which originates from destructive quantum interference, has become one of the pillars on which our understanding of electron transport in disordered systems rests. It similarly has played an important role in the description of the integer quantum Hall effect. More recently it has had far-reaching implications for non-equilibrium physics leading to the discovery of many-body localization (MBL) \cite{Basko2006,Gornyi2005}. Localization is thus the paradigmatic ingredient for avoiding thermalization in many-body quantum systems~\cite{Nandkishore2015,Alet2018,Abanin_review}.

Basically since its original discovery, there have been efforts to establish the possibility of localization due to quantum interference in many-body systems \textit{without} quenched disorder. Prime candidates of translationally invariant systems displaying localization were mixtures of heavy and light particles~\cite{kagan1984localization}, which however only show transient localization behavior before eventual thermalization~\cite{YaoQuasiMBL}. Despite much effort, only recently was genuine \textit{disorder-free localization} (DFL) demonstrated in a concrete and surprisingly elementary microscopic model~\cite{Smith2017}.  The original proposal featured an integrable model with a local $\mathbb{Z}_2$ symmetry which allowed an exact mapping to an extensive collection of Anderson insulators with binary disorder. Crucially, the Hamiltonian and initial states of quench protocols have full translation invariance but the latter are an extensive superposition of individual binary disorder configurations, i.e., DFL exploits a form of quantum parallelism~\cite{paredes2005exploiting}. A remarkable consequence of this is the dephasing between an extensive number of many-body sectors leading to a tunable growth of the entanglement entropy to a volume-law regime. This happens despite the fact that  particle transport and correlation spreading are absent in DFL, akin to the MBL phenomenology~\cite{smith2017absence}.

The discovery of DFL generated a lot of research activity and it has become clear that the DFL phenomenon is general and does neither rely on integrability nor any mapping to Anderson insulators~\cite{smith2017absence}. In fact, the key ingredient for DFL is the presence of \emph{local} symmetries. This has lead to  connections to the research field of lattice gauge theories (LGTs)~\cite{Brenes2018}. Many examples of DFL in models of strongly coupled LGTs have been explored~\cite{Smith2018,papaefstathiou2020disorder,smith2019logarithmic,HalimehStabilizingDFL,chakraborty2022disorder} and recent work proposed that the DFL mechanism might be relevant in metallic materials with nonlinear electron-phonon coupling~\cite{sous2021phonon}. While initial work focused on one-dimensional systems, current efforts aim at extending DFL to higher dimensions. However, progress is hampered by the limitation of numerical methods for studying strongly coupled LGTs and for calculating the time dependence of interacting quantum many-body systems. 

Despite copious theoretical attention to DFL including experimental proposals \cite{HalimehEnhancingDFL,Homeier2023realistic}, efforts of experimental realization have commenced only very recently~\cite{Gyawali2024}.
The key technical challenge has been the implementation of three-qubit interactions obeying the required local $\mathbb{Z}_2$ symmetries. 
The same building blocks enter LGT implementations, e.g., those realized  in analog~\cite{yang2020observation} and digital~\cite{martinez2016real,Mildenberger2022} quantum simulators, but inherent errors and noise limit attainable system sizes and times. The recent experiment~\cite{Gyawali2024} on DFL has been enabled by a modification of the original model that renders it nonintegrable. Such a model is useful when it comes to probing localization in interacting systems. However, studying an integrable model offers the unique advantage of rigorously benchmarking a quantum computer. Typical NISQ experiments are limited to  short evolution times and system sizes in that they can still be more or less reproduced on classical machines. However, as the technology advances, longer evolution times and larger system sizes will become accessible on quantum computers, matching which classical devices are expected to find increasingly challenging. These experiments will then require reliable benchmarking: integrable models then become a tool of choice for this purpose, as they allow access to arbitrary system sizes and evolution times with classical (numerical or analytical) means.

Here, we present a  theoretical proposal for realizing a tunable model of DFL ideally suited for exploring near-term NISQ capabilities. Our proposal  has the following desirable features. First, it starts form a soluble point, thus allowing quantitative benchmarking against classical computers.  Second, deformations readily turn this model into a genuinely hard-to-realize quantum many-body problem--but retain a clear and predictable phenomenology. Third, a tuning parameter in the model allows for the adjustment of localization time and length scales, thus permitting placing different regimes in the experimental focus. Fourth, there exist observables which are not self-averaging, and hence beyond the purview of present classical computers for all but the most moderate systems sizes. Fifth, we present an interesting phenomenology involving Renyi entropies to probe the characteristic specifics of the DFL phenomenology. {Sixth, none of the above are intrinsically linked to low dimension, and we mention generalisation to higher dimension.}

In the remainder of this paper, we first introduce the model and present its central properties and novel experimental probes/signatures. We then present its circuit implementation and analyze the role of noise alongside possible error mitigation. We conclude with a summary and discussion. 

\section{Microscopic model with local $\mathbb{Z}_2$ symmetry}
We propose to consider the basic original model that demonstrates the physics of DFL~\cite{Smith2017}. The advantage is that it has an exactly soluble limit in which some diagnostics of DFL can be efficiently calculated. This enables to provide benchmark results for large system sizes and late times in regimes which are genuinely hard to simulate. Moreover, the system can easily be tuned into a non-integrable regime whose quantitative study is beyond current numerical capabilities. 

We first focus on a one-dimensional chain with spinless fermions $\hat{f}_j$ living on the sites, coupled to spin-$1/2$ degrees of freedom, $\hat{\sigma}_{j,j+1}$, positioned on the bonds. It is described by the Hamiltonian
\be \label{hamilt}
\begin{split}
\hat{H}_0 =&-J\sum_{j}\big(\hat{\sigma}^z_{j, j+1}\hat{f}^{\dagger}_j\hat{f}_{j+1} +\hc\big)\\
&+h\sum_j \hat{\sigma}^{x}_{j-1,j} \hat{\sigma}^{x}_{j,j+1},
\end{split}
\ee
where $h$ is the strength of spin couplings, and the fermions and spin operators obey the standard (anti)commutation relations.

A defining feature of the Hamiltonian \eqref{hamilt} is that it has local $\mathbb{Z}_2$ symmetry: $\hat{f}_j \to \theta_j \hat{f}_j$ and $\hat{\sigma}^z_{j,j+1} \to \theta_j \hat{\sigma}^z_{j,j+1} \theta_{j+1}$ with $\theta=\pm 1$. The underlying physics is simply that a local fermion parity change is linked to a change of the bond spin (in the $x$-polarized basis), which results in an extensive number of conserved quantities (charges) $\hat{q}_j = (-1)^{n_j}\hat{\sigma}^{x}_{j} \hat{\sigma}^{x}_{j+1}$, where $\hat{n}_j = \hat{f}^{\dagger}_j\hat{f}_{j}$ is the fermionic number operator on site $j$. These so-called charges, which commute with the Hamiltonian on account of the local symmetry, take on values $\pm 1$. We note that in contrast to the standard description of the corresponding $\mathbb{Z}_2$ LGT, we do not work within a single fixed gauge-invariant sector. Since we treat both the fermions and bond spins as physical degrees of freedom, DFL allows for the superposition of different charge sectors. 

One can add more general terms to the Hamiltonian, e.g., a simple transverse field, which breaks integrability of the model and spoils any mapping to a short-range interacting binary disorder model ~\cite{Smith2018}. It is then directly linked to standard $\mathbb{Z}_2$ LGTs such as variants of the famous Schwinger model of $(1+1)$-dimensional QED~\cite{Brenes2018,Smith2017}. 
Here, in order to elucidate the DFL mechanism, we restrict the discussion to the above-mentioned model, and stress that DFL is a robust feature as long as the $\mathbb{Z}_2$ symmetry is preserved~\cite{Smith2018,HalimehStabilizingDFL}. In this context, we note that DFL is a general phenomenon and has also been established in quantum link models and LGTs with variants of $U(1)$ symmetry~\cite{Brenes2018,papaefstathiou2020disorder,karpov2021disorder,chakraborty2022disorder,osborne2023disorderfree}.

To first elucidate the appearance of DFL, it is instructive to perform a duality transformation for the spins, similar to that known for the Ising model~\cite{Kramers1941}. In the subspace fixed by a particular charge configuration (i.e., a given string of $\pm1$), the Hamiltonian \eqref{hamilt} maps to a tight-binding model with a binary potential corresponding to this string~\cite{Smith2017}:
\be \label{hamilt_bin}
\hat{H}_{\{q_j\}} {=} {-}J\sum_{j}\big(\hat{c}^{\dagger}_{j}\hat{c}_{j+1} {+} \hc \big) {+} h\sum_{j}q_j \big(2\hat{c}^{\dagger}_{j}\hat{c}_{j} {-} 1\big).
\ee
The Hamiltonian \eqref{hamilt_bin} provides insight into the mechanism of localization in a translationally invariant system~\eqref{hamilt}, namely, that conserved charges can effectively play the role of a random potential. 
To corroborate this statement with a specific example, let us consider an initial state of the system in the Hilbert space of~\eqref{hamilt} with all bond spins polarized along the $z$-axis. Here we introduce the following notation, $\ket{\uparrow}$ refers to a $z$-polarized spin, and $\ket{\rightarrow}$ refers to an $x$-polarized spin. Noting that $\ket{\uparrow}= \big[\ket{\rightarrow} + \ket{\leftarrow}\big]/\sqrt{2}$, one can see that a fully $z$-polarized spin-state corresponds to an equal-weight superposition of all charge configurations
\be
\ket{\uparrow \uparrow \ldots\uparrow }_{\sigma} \otimes \ket{\psi_{f}} = \frac{1}{2^{N}}\sum_{\{q_j\} = \pm 1}\ket{q_1 q_2 \ldots q_{N}}\otimes\ket{\psi_{f}},
\label{EqInitState}
\ee
where $\ket{\psi_{f}}$ is the fermionic part of the wave function which turns out to be the same for $\hat c$ and $\hat f$ fermions. 
From the duality transformation, it is then transparent that localization occurs in the case of $z$-polarized initial states. Note that both the Hamiltonian and initial states are translationally invariant as a hallmark of DFL, but it is the quantum superposition of an extensive number of disordered charge sectors, each typical one with a disordered potential, that leads to localization. 

On the other hand, let us consider initial states with all spins polarized along the $x$-axis, which correspond to a single fixed configuration of charges. If bond-spins are aligned in such a way that all charges have the same value (either $+1$ or $-1$), localization behavior is absent. Note that only for the fine-tuned Hamiltonian~\eqref{hamilt} does DFL directly map onto a binary disorder problem of {\it free} fermions, but the localization persists (and generically becomes of the MBL-type) for more general interactions obeying the local $\mathbb{Z}_2$ symmetry.

The strong initial state dependence is an ideal target for diagnosing DFL with quantum simulators. In the following , we are going to observe (non-)ergodicity by studying the dynamical response of the system following a quantum quench from a (``localized'') ``delocalized'' initial state, e.g., one with all bond-spins polarized along the $z$-axis, and the second with $x$-polarized spins. The fermion subsystem is prepared in a convenient charge domain wall state, e.g., $\ket{11110000}$ for a chain of length $L = 8$, but one can consider other initial states described by any Slater determinant. 

\begin{figure}[h]
\includegraphics[width=8.5cm]{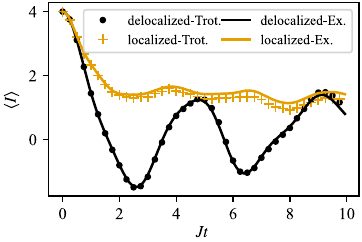}
\caption{Fermion imbalance as a probe of localization for a chain of $L=8$ sites, $J = 1$, $h = 1$ and Floquet step $T=0.25$, following a quench from a domain-wall initial state for the fermions. Solid lines are continuous time evolution results obtained via exact diagonalization (ED) and discrete data points are obtained from exact quantum circuit implementations of the Floquet time evolution.} \label{rho}
\end{figure}

\section{Measures of DFL}
In order to diagnose the DFL physics we can make use of different measurements on the quantum hardware. 

\subsection{Conventional probe of initial state memory}

First, we use one of the simplest measures of localization:  the long-time memory of the initial state after a quantum quench. In order to probe this memory, we measure the imbalance
\be\label{dens_imb}
I(t) = \sum_{j = 1}^L\big(2\langle\hat{n}_j(0)\rangle-1\big)\langle\hat{n}_j(t)\rangle.
\ee
The same measure was used, for example, to identify MBL in cold-atom experiments~\cite{Schreiber2015}. For a generic ergodic system, $I(t)$ vanishes at long times (for large systems). For localized non-ergodic systems, one instead finds saturation to a finite asymptotic value, demonstrating persistent memory of the initial state.

In Figure~\ref{rho}, we show that already for small systems   one can observe the distinct behavior albeit with the expected finite size effects (the delocalized state oscillates around zero and revivals occur for later times). The upshot is that DFL in our model is clearly observable for small system sizes and reasonable time scales. 

A key advantage is that in the non-interacting model we have full control over the localization length $\xi$, which is set by the transverse coupling $h$, see Fig.~3 of Ref.~\cite{Smith2017}. Already for $h\approx J$ the system is strongly localized with $\xi \approx 1$ and the long-time value of the imbalance is large. Reducing $h/J$ increases $\xi$ which requires the simulation of larger systems $L> \xi$ to observe localization. The upshot is that the soluble points can be used for benchmarking the size dependence of quantum simulators.

\begin{figure}[h]
\includegraphics[width=8.5cm]{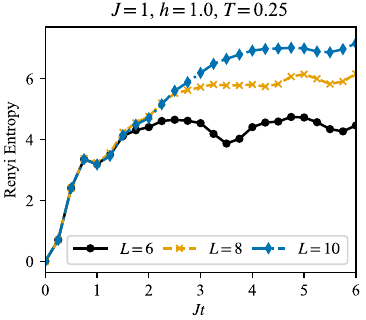}
\caption{The half-chain second-order Renyi entropy for chains of $L=8$, $10$ and $12$ sites. $J = 1$, $h = 1$, and Floquet step $T=0.25$, following a quench from a domain-wall initial state for the fermions. After a first area-law regime the steady state value grows linearly with system size  characteristic of the DFL phenomenon.} \label{entropy}
\end{figure}

\subsection{New probes of quantum parallelism}

Second, we can exploit the unique capabilities of digital quantum computers to probe DFL characteristics which are hard to obtain in analogue quantum simulators. Such a measure of localization is the half chain entanglement entropy. Concretely, we focus on the Renyi entropy which can be efficiently extracted via randomized measurements~\cite{Brydges2019} and has been previously implemented~\cite{Vovrosh2021}. Crucially, it not only provides an additional diagnostic of DFL, but in fact is able to diagnose its key feature.
In Figure~\ref{entropy}, we show benchmark results for the half-chain Renyi entropy starting from a localized initial state. Remarkably, one can observe that despite the localization of particle transport the entropy reaches a volume-law regime. The reason is that the entropy not only probes the localization physics within each charge sector but also the dephasing {\it between} different sectors in the extensive superposition, see Eq.~\eqref{EqInitState}. Hence, it directly confirms the coherence between the charge sectors~\cite{smith2017absence}. Therefore, we propose to use randomized measurements to directly confirm the key signatures of quantum parallelism. 

Again, we can use knowledge of the soluble limit to benchmark the size dependence. For short times, the Renyi entropy shows an area law, which after a time scale $\tau$ crosses over to the volume law, for example $\tau\approx 2.5/J$ in Fig.~\ref{entropy}. The time scale $\tau$ increases as $(h/J)^2$, see inset of Fig.~1 in Ref.~\cite{smith2017absence}. We find that for larger localization (small $\xi$) one needs to simulate longer times to observe the volume law scaling, or vice versa, to observe volume law already after short times we need large system sizes (to observe localization at large $\xi$).

Third, one could measure the Green's function for the fermionic degrees of freedom
\be\label{GF}
G_{jk}(t) = \langle \hat f_j^{\dagger}(t) \hat f_k (0)\rangle.
\ee
which provides yet another measure of localization (not shown). In contrast to simple density averages the quantity is not self-averaging, meaning that one needs to sum over all charge configurations in Eq.~\eqref{EqInitState}. Thus, even for the free fermion soluble model certain correlators are hard to compute with classical numerics. Finally, one can also probe localization and the special phase coherence between the different sectors via out-of-time order correlators~\cite{smith2019logarithmic}.

\section{Quantum circuit for Floquet time evolution}
In this section, we elaborate on how our DFL model can be implemented in detail for Floquet time evolution. Note, the DFL mechanism works both for continuous time evolution and large Floquet periods but the latter requires much fewer resources.  We provide benchmark results for small system sizes comparable to current capabilities~\cite{Smith2019}, which can then be scaled up to probe the system size and time scales which can only be compared to the soluble point.

First, spinless fermions on the sites can be mapped to spins-$1/2$ via the Jordan--Wigner transformation~\cite{JW}. As a result, instead of a system containing both fermions and spins, we consider a pure spin system. Correspondingly, the initial system of $N$ sites with fermions on some of them and $N$ bond spins can be simulated via $2N$ qubits. It is worth noting that the boundary conditions of the spin problem depend on the parity of the total number of fermions: for even total number of fermions, periodic boundary conditions (PBC) transform into anti-periodic boundary conditions (APBC); and in the case of odd  total number of fermions, PBC transform into PBC. The Hamiltonian \eqref{hamilt} takes the following form (here $\hat{S}$ refers to spins on the sites):
\be \label{hamilt_spin}
\begin{split}
\hat{H} =& -J\sum_{j}\big(\hat{\sigma}^{z}_{j,j+1}\hat{S}_{j}^{+}\hat{S}_{j+1}^{-} + \hc\big)\\
&+ h\sum_j \hat{\sigma}^{x}_{j-1,j} \hat{\sigma}^{x}_{j,j+1}.
\end{split}
\ee
The initial state of the fermionic subsystem maps to a domain wall of ``up'' and ``down'' spins: $\ket{\psi_f} = \ket{\uparrow \uparrow \ldots \downarrow \downarrow}$. 

Second, we use Floquet dynamics by splitting the time-evolution operator for $n$ steps into a sequence of discrete operators: $\hat{U}(nT) = \prod_{i = 1}^{n}\hat{U}(T)$ with fixed time-period $T$. Each application of $\hat{U}(T) = e^{-i\hat{H}T}$ is called a Floquet step. 

\subsection{Circuit implementation}
Now the Floquet time-evolution operator can be decomposed by defining the following operators:
\be
\begin{split}
&\hat{A}_j = e^{iJT\hat{\sigma}^{z}_{j,j+1}(\hat{S}_{j}^{+}\hat{S}_{j+1}^{-} + \hc)};\\
&\hat{B}_j = e^{-i hT\hat{\sigma}^{x}_{j-1,j} \hat{\sigma}^{x}_{j,j+1}},
\end{split}
\ee
which then give the full evolution over one period:
\begin{equation}
\hat{U}(T) = \prod_{j~\text{odd}}\hat{A}_{j} \prod_{j~\text{even}}\hat{A}_{j}
\prod_{j~\text{odd}}\hat{B}_j\prod_{j~\text{even}}\hat{B}_j.
\end{equation}
The operators $\hat{A}_j$ and $\hat{B}_j$ then need to be decomposed into sequences of single-qubit gates and CNOTs (or iSWAPS, see Appendix~\ref{app}) readily available on current quantum computers. This will depend on the available gate set, the geometry of the quantum computer, and the chosen embedding of the model. In the following, we assume access to arbitrary single qubit gates as well as CNOT gates (which could equivalently be replaced by CZ with the appropriate addition of Hadamard gates). We also assume that we have $1+1$D nearest-neighbor connectivity (which could be part of another $2+1$D geometry), and that we embed our model with alternating $S$ and $\sigma$ spin species.\\

We employ the following single-qubit gates:

-- Pauli matrices:
\be
\begin{split}
\Qcircuit @C=.5em @R = .5em{
& \gate{X} & \qw
} &= 
\begin{pmatrix}
0 & 1\\
1 & 0\\
\end{pmatrix},~~\Qcircuit @C=.5em @R = .5em{
 & \gate{Y} & \qw
} = 
\begin{pmatrix}
0 & -i\\
i & 0\\
\end{pmatrix}\\
\Qcircuit @C=.5em @R = .5em{
 & \gate{Z} & \qw
} &= 
\begin{pmatrix}
1 & 0\\
0 & -1\\
\end{pmatrix},
\end{split}
\ee

-- Phase gates:
\be
\Qcircuit @C=.5em @R = .5em{
& \gate{H} & \qw
} = \frac{1}{\sqrt{2}}
\begin{pmatrix}
1 & 1\\
1 & -1\\
\end{pmatrix},~~
\Qcircuit @C=.5em @R = .5em{
 & \gate{S} & \qw
} =
\begin{pmatrix}
1 & 0\\
0 & i\\
\end{pmatrix},
\ee
(the Hadamard and the S-phase gate)

-- Rotation gates:
\begingroup
\renewcommand{\arraystretch}{1.5}
\begin{equation}
\begin{split}
\Qcircuit @C=.5em @R = .5em{
& \gate{R_{X}(\theta)} & \qw} 
&= 
\begin{pmatrix}
\cos\frac{\theta}{2} & -i\sin\frac{\theta}{2}\\
-i\sin\frac{\theta}{2} & \cos\frac{\theta}{2}\\
\end{pmatrix}\\
\Qcircuit @C=.5em @R = .5em{
 & \gate{R_{Y}(\theta)} & \qw
} &= 
\begin{pmatrix}
 \cos\frac{\theta}{2} & -\sin\frac{\theta}{2}\\
 \sin\frac{\theta}{2} & \cos\frac{\theta}{2}\\   
\end{pmatrix}\\
\Qcircuit @C=.5em @R = .5em{
 & \gate{R_{Z}(\theta)} & \qw
} &= 
\begin{pmatrix}
 e^{-i\frac{\theta}{2}} & 0\\
 0 &  e^{i\frac{\theta}{2}}\\   
\end{pmatrix}.
\end{split}
\end{equation}
\endgroup

Starting with the $\hat{B}_j$, we can use the well known decomposition into standard gate sets, namely
\vspace{0.1cm}
\be
\Qcircuit @C=.5em @R = 0.8em{
\lstick{\sigma_{j-1,j}} & \gate{H} & \ctrl{1} & \qw & \ctrl{1} & \gate{H} & \qw \\
\lstick{\sigma_{j,j+1}} & \gate{H} & \targ & \gate{R_Z(2hT)} & \targ & \gate{H} & \qw 
}
\ee
\vspace{0.1cm}
However, when embedding the model with alternating $S$ and $\sigma$ species, this requires long-range CNOT gates, which cannot be implemented directly. Instead of using swap operations to fix this, we can make use of partial swaps to get the nearest-neighbor decomposition
\vspace{0.1cm}
\be
\Qcircuit @C=.4em @R = 0.8em{
\lstick{\sigma_{j-1,j}} & \gate{H} &\ctrl{1} & \qw & \ctrl{1} & \qw & \qw & \ctrl{1} & \qw & \ctrl{1} & \qw & \gate{H} & \qw \\
\lstick{S_j} & \qw & \targ &  \ctrl{1} & \targ & \ctrl{1} & \qw & \targ & \ctrl{1} & \targ & \ctrl{1} & \qw & \qw \\
\lstick{\sigma_{j,j+1}} & \gate{H} & \qw & \targ & \qw & \targ & \gate{R_Z(2hT)} & \qw & \targ & \qw & \targ & \gate{H} & \qw 
} \!
\ee
\vspace{0.1cm}
Finally, the $\hat{A}_j$ operators can be implemented with the following decomposition

\vspace{0.1cm}
\be
 \Qcircuit @C=.5em @R = 1em{
&\lstick{S_j} & \multigate{2}{C} & \targ & \qw & \gate{R_Y(+JT/2)} & \qw & \targ & \multigate{2}{D} & \qw \\
&\lstick{\sigma_{j,j+1}} & \ghost{C} & \ctrl{-1} & \ctrl{1} &\qw & \ctrl{1} & \ctrl{-1} & \ghost{D} & \qw\\
&\lstick{S_{j+1}} & \ghost{C} & \qw & \targ & \gate{R_Y(-JT/2)} & \targ & \qw & \ghost{D} & \qw
    }
\ee
where blocks $C$ and $D$ denote the circuits
\vspace{0.1cm}
\be
 \Qcircuit @C=.5em @R = 1em{
&\lstick{S_j} & \gate{H} & \gate{S} & \gate{H} & \ctrl{1} & \qw & \ctrl{1} & \qw &  \qw & \qw & \gate{S} & \gate{H} & \qw \\
&\lstick{\sigma_{j,j+1}} & \qw & \qw & \qw & \targ & \ctrl{1} & \targ & \ctrl{1} & \qw & \qw & \qw & \qw & \qw\\
&\lstick{S_{j+1}} & \gate{H} & \gate{S} & \gate{H} & \qw & \targ & \qw & \targ  & \gate{H} & \gate{Z} & \gate{S} & \gate{H} & \qw
    }
\ee
and
\vspace{0.1cm}
\be
 \Qcircuit @C=.5em @R = 1em{
&\lstick{S_j} & \gate{H} & \gate{Z} & \gate{S}  & \ctrl{1} & \qw & \ctrl{1} & \qw & \gate{H} & \gate{Z} & \gate{S} & \gate{H} \qw \\
&\lstick{\sigma_{j,j+1}} & \qw & \qw & \qw & \targ & \ctrl{1} & \targ & \ctrl{1} & \qw & \qw & \qw & \qw & \qw \\
&\lstick{S_{j+1}} & \gate{H} & \gate{S} & \gate{H} & \qw & \targ & \qw & \targ  & \gate{H} & \gate{Z} & \gate{S} & \gate{H} & \qw
    }
\ee
correspondingly, which respects the nearest-neighbor connectivity. In the full Trotter decomposition, we are able to apply some CZ gates in parallel, which allows us to reduce the two-qubit gate depth by 4 (except in the final step). In total, we require $20$ CNOT/CZ layers per Floquet period, where we have assumed that interleaving single qubit gates is relatively low cost and can be neglected.

So far, we have implemented the quantum circuit for the three-qubit gates via the available standard two-qubit gates. However, a direct analog implementation of the three-qubit gates is in principle possible. In addition to the superconducting qubit implementation proposed here, trapped ion quantum simulators also allow for a direct implementation of three-qubit gates~\cite{katz2022n}, therefore, providing an alternative platform for realizing DFL. 

We have implemented the Floquet quantum circuit  and representative results of a noise-free simulation are shown in Fig.~\ref{rho} (points) for Floquet period $T=0.25$ (all units in terms of $J$). The distinction between localized and delocalized initial states are clearly observable.

\section{Noise, imperfections, and feasibility}
The presence of errors and noise is the biggest challenge for reliably observing DFL (or any form of quantum non-ergodicity) on quantum computers. For example, in the recent work on realizing DFL a different simpler three-qubit interaction was employed corresponding to a smaller number of CNOTs~\cite{Gyawali2024}. 
In this section, we shortly outline several error mitigation strategies and then comment on the scalability for benchmarking.

First of all, one can notice that the Hamiltonian Eq.~\eqref{hamilt} conserves the number of fermions, which, after the Jordan--Wigner transformation, maps to the conservation of total magnetization of spins on the sites $\sum_j S^z_j$. A simple but powerful mitigation is to disregard any measurements of states violating the conservation law~\cite{Smith2019}.

The presence of noise and decoherence leads to a decay of an ideally persistent signal as shown for example in Google's realization of eigenstate order (or discrete time crystal behavior)~\cite{Mi2022}. The authors of this previous work introduced a quantity aimed at measuring the accuracy of the implementation of the unitary time evolution: $N(t) = |\langle \hat{U}(t)\hat{U}^{-1}(t)\rangle|$. The idea is closely related to the concept of a Loschmidt echo~\cite{Loschmidt}. In an ideal machine, this quantity is of course always equal to $1$. In real systems, however, noise and decoherence results in a decrease of this value as a function of time. One can then divide measurement results by $N(t)$ to restore the noise-free behavior proving non-ergodicity, e.g., see Fig.~2c in Ref.~\cite{Mi2022}.

We have implemented this echo-based error mitigation with Qiskit's simulator with noise, but, rather surprisingly, it fails. The reason turns out to be interesting and related to the DFL mechanism itself: since only the fermionic subsector is localized (the original bond spins are ergodic~\cite{Smith2017}), the full Loschmidt echo incorporating the decay of both sectors decays too fast.

\begin{figure}[]
\includegraphics[width=0.91\linewidth]{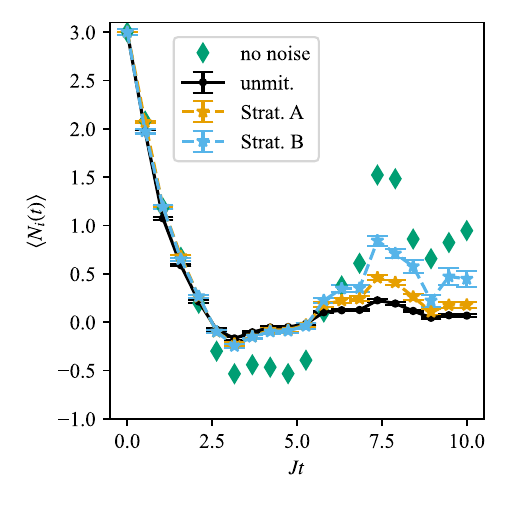}
\includegraphics[width=0.91\linewidth]{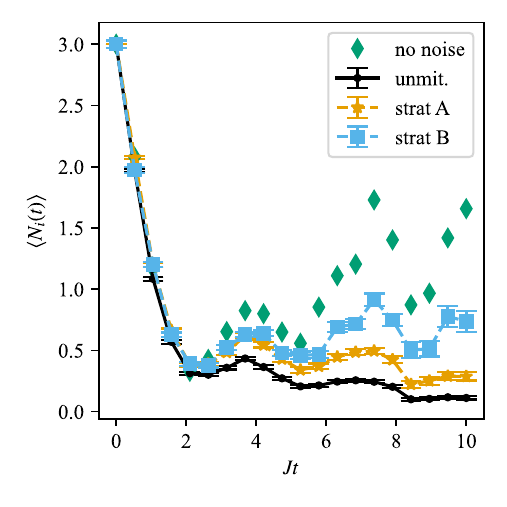}
\caption{The average imbalance computed with the Floquet quantum circuit implemented with Qiskit's QasmSimulator for a chain of $L=6$ sites, Floquet period $T = 0.5$, $J = 1$, and $h = 1$. Upper panel shows results for a delocalized initial state and the lower pane for a localized one. The green symbols are for an ideal implementation of the quantum circuit. The solid points are for a simulator with a realistic noise strength (0.005). The dashed curves, the green one and the lime one, represent the results from the latter noisy simulator after employing the error mitigation strategies described in the main text (strategies A and B correspondingly). The persistence of the imbalance can be recovered.} \label{noise}
\end{figure}

We have developed and tested two different error mitigation strategies which directly take into account the fact that the LGT Hamiltonian~\eqref{hamilt} splits into different charge sectors with fermion and bond spin degrees of freedom. Noise couples differently to the different sectors but only the fermionic matter sector shows localization. In the following, we explain a novel subsector-specific error mitigation strategy. 

To implement strategy A, we employ that the actually implemented quantum circuit only weakly depends on the parameter $T$ (the Floquet period). We can then measure the decay of a local operator, e.g., only for the fermionic sector of $\hat{\sigma}$ spins, $\tilde N(t) =\langle \sigma_i \rangle$ with $t=n T$ for $n$ steps. We consider the Floquet time evolution of this value, keeping the number of periods  constant (and consequently, the depth of the corresponding quantum circuit persists), but decreasing the size of the period T. In an ideal system with $T$ going to zero, the studied quantity tends to its initial state value. In real circuits with noise, the ratio of the measured value and its value measured in the initial state decreases for increasing number of Trotter steps. The decay eventually reaches a steady value and becomes independent of $T$. We can then use $\tilde N(t)$ to mitigate local measurement results as above. 

Strategy B is a modification of the aforementioned Loschmidt echo-based error mitigation strategy. Instead of measuring the full Loschmidt echo, we perform forward and backward evolution and measure the average values of local operators: $\tilde{N}_i(t) = \langle \tilde{\psi}(t)|\hat{S}^z_i|\tilde{\psi}(t)\rangle$, where $|\tilde{\psi}(t)\rangle = \hat{U}(-t/2)\hat{U}(t/2)|i\rangle$ ($|i\rangle$ is the initial state). Note that we evolve the initial state up to $t/2$ in order to preserve the total number of Floquet steps.

In Figure~\ref{noise}, we compare the ideal quantum circuit implementation (green) compared to a noisy one (black). Note, because the circuit with noise is computationally more demanding the system sizes are smaller. One can observe that the noise induced decay destroys the persistence of the initial state density imbalance. However, this artificial decay can be removed via dividing out the local error $\tilde N(t)$ as obtained from our strategies A and B. Hence, our DFL specific error mitigation strategies indeed approaches the noise-free result and the artificial decay of the persistent imbalance (lower panel starting from a localized initial state) can be removed.

\section{DFL in $2+1$ dimensions}
DFL is not restricted to $1+1$D but can similarly appear in $2+1$D. Again, only for a special free-fermion soluble model~\cite{Smith2018} is the quantitative behavior known, but for generic systems classical numerical methods are limited to very small system sizes. This is especially a problem since typical localization lengths of disordered systems of the Anderson type have much larger localization lengths compared to $1+1$D. Hence, just like in $1+1$D having a soluble limit will be an ideal testbed for pushing quantum computers. The $2+1$D generalization of Eq.~\eqref{hamilt} is a $\mathbb{Z}_2$ LGT in $2+1$D described by the Hamiltonian
\be \label{hamilt_spin2D}
\begin{split}
\hat{H}_{2+1\mathrm{D}} =& -J\sum_{\langle j k \rangle}( \hat{S}_{j}^{+} \hat{\sigma}^{z}_{j,k}\hat{S}_{k}^{-} + \hc)\\
&+ h_{\perp}\sum_{\langle j k \rangle} \hat{\sigma}^{x}_{j,k} + h \sum_{r^\ast} \prod_{\langle j k \rangle \in r^\ast} \hat{\sigma}^{z}_{j,k}.
\end{split}
\ee
with the last term the ``magnetic-field energy'' given by the operator of four-bond spins around the plaquette labeled by $r^\ast$. For $h_{\perp}=0$ the duality transformations of the $2+1$D Ising gauge theory can be used to show that it again maps to a system of free fermions in a discrete disorder potential. Similarly to the $1+1$D case, the localization length is a tunable function of $K$, see Fig.~7 of Ref.~\cite{Smith2018}, which can be used to benchmark future large quantum computers probing $2+1$D physics. The time scale when the entanglement crosses over from area to volume law scales as $\tau\propto (h/J)^2$.   A nonzero $h_{\perp}$ preserves the local symmetry but turns it into a generic interacting model also expected to display DFL which is beyond the reach of any classical numerical method.

\section{Summary and Discussion}
We have introduced the quantum circuit implementation and benchmark results for a special model displaying the DFL phenomenon. The key advantage of our system is the availability of a free-fermion soluble point. While we have shown that currently available hardware, e.g.  realistic values of qubit numbers and circuit depths, could implement it in principle already now, the current level of noise is a real challenge. Therefore, we have also introduced error mitigation strategies exploiting that the LGT Hamiltonians of DFL split into multiple components.

One of the key goals in the field of quantum computation/simulation is to have a reliable way of showing that quantum computers can tackle genuinely hard quantum many-body problems. We argue that our DFL proposal has several advantages in this context: (i) The soluble limit can be used to benchmark the implementation, namely the degree of localization (the value of the localized plateau in Fig.~\ref{rho}) depends on the intrinsic localization length $\xi$ which decreases for increasing $h/J$. At the same time, the time scale $\tau$ where the entropy crosses over to volume law behavior increases as $(h/J)^2$, again a tunable behavior ideally suitable for benchmarking.  (ii) Already for the free fermion soluble limit some of the correlators, e.g. the Greens function Eq.~\eqref{GF}, are not self averaging and thus cannot be calculated on classical computers for larger system sizes; (iii) The simple addition of a transverse field breaks the integrability of the model and turns it into a numerically intractable problem for classical computers, while DFL remains robustly present; (iv) The phenomenon of DFL is not restricted to one spatial dimension, but can be straightforwardly generalized to $2+1$D~\cite{Smith2018}, where it would provide an even more challenging benchmark for future generations of quantum computers.

Overall, in conjunction with a reliable error mitigation strategy the observation of quantum non-ergodicity of the DFL type would establish quantum parallelism on available quantum computers and benchmark their current and future capabilities. An implementation of our model will rely on the efficient encoding of three-qubit gates, the observation of localization of correlations in conjunction with a tunable growth of the entanglement (or Renyi) entropy to a volume law regime. A realisation of DFL then not only provides confirmation of the quantum parallelism underpinning this 
complex quantum many-body interference effect from local symmetries, but, at the same time, it can be used as a non-trivial tool for benchmarking the capabilities of quantum computers in regimes with massive many-body entanglement. 

\begin{acknowledgments}
    The authors are grateful to Erik Gustafson and Adam Smith for fruitful discussions and input on the circuit implementation, and to Ian P.~McCulloch and Jesse J.~Osborne for fruitful discussions on DFL in $2+1$D. J.C.H.~acknowledges support from the Max Planck Society and the Emmy Noether Programme of the German Research Foundation (DFG) under grant no.~HA 8206/1-1. This work was supported by the Deutsche Forschungsgemeinschaft via Research Unit FOR 5522 (project-id 499180199), as well as the cluster of excellence ct.qmat (EXC-2147, project-id 390858490). J.K.~acknowledges support from the Deutsche Forschungsgemeinschaft (DFG, German Re- search Foundation) under Germany’s Excellence Strat- egy–EXC– 2111–390814868, DFG grants No.~KN1254/1- 2, KN1254/2- 1, and TRR 360 - 492547816; as well as the Munich Quantum Valley, which is supported by the Bavarian state government with funds from the Hightech Agenda Bayern Plus.
\end{acknowledgments}

\appendix
\section{Sycamore architecture}\label{app}
All circuits in this paper are implemented with CNOT entangling gates readily available on IBM processors. The corresponding implementation for the Google Sycamore architecture is easily adapted, e.g., the circuit for the $3$-qubit interaction is given by~\cite{Mildenberger2022}:
\vspace{0.5cm}
\begin{widetext}
\begin{align}
\Qcircuit @C=.5em @R = 1em{
    \lstick{\sigma_i} & \multigate{1}{\sqrt{iS}} & \multigate{1}{\sqrt{iS}}
    & \qw & \qw & \qw  &  \multigate{1}{\sqrt{iS}^{\dagger}} & \multigate{1}{\sqrt{iS}^{\dagger}} &\qw\\
    \lstick{\tau_{i,i+1}} & \ghost{\sqrt{iS}} & \ghost{\sqrt{iS}} 
    & \multigate{1}{\sqrt{iS}} & \gate{R_z(-J\Delta t)} & \multigate{1}{\sqrt{iS}^{\dagger}} & \ghost{\sqrt{iS}^{\dagger}} & \ghost{\sqrt{iS}^{\dagger}} & \qw \\
    \lstick{\sigma_{i+1}} & \qw & \qw & \ghost{\sqrt{iS}} & \gate{R_z(+J\Delta t)} & \ghost{\sqrt{iS}^{\dagger}} & \qw & \qw & \qw 
}
\end{align}
\end{widetext}

\bibliography{biblio}

\end{document}